\documentclass[a4paper,11pt]{article}
\pdfoutput=1 

\usepackage{jheppub}     
\usepackage[T1]{fontenc}  
\usepackage{bm}
\usepackage{mathrsfs}
\usepackage{cancel}
\usepackage[normalem]{ulem}
\usepackage{slashed, braket, dirtytalk}
\usepackage{xcolor}
\usepackage{float, soul, parskip, lmodern}
\usepackage{caption}
\usepackage{siunitx}
\usepackage{subcaption}
\usepackage{gensymb}
\usepackage{natbib}
\usepackage{bbold}
\usepackage{booktabs}

\def\be{\begin{equation}}
\def\ee{\end{equation}}

\newcommand{\bear}{\begin{eqnarray}}
\newcommand{\bea}{\begin{eqnarray}}
\newcommand{\eear}{\end{eqnarray}}
\newcommand{\eea}{\end{eqnarray}}
\newcommand{\ba}{\begin{array}}
\newcommand{\ea}{\end{array}}
\newcommand{\bi}{\begin{itemize}}
\newcommand{\ei}{\end{itemize}}

\def\a{\alpha}		\def\b{\beta}

\def\s{\sigma}

\renewcommand{\b}[1]{\textbf{#1}}

\def\II{\relax{\rm I\kern-.18em I}}

\def\s{\sigma}

\def\a{\alpha}
\def\b{\beta}

\def\<{\big\langle}
\def\>{\big\rangle}

\def\II{{\cal I}}

\def\nn{\nonumber}
\newcommand{\red}[1]{{\color{red} #1 \color{black}}}

\newcommand{\teal}[1]{{\color{teal} #1 \color{black}}}

\newcommand{\ov}{\overline}

\newcommand{\PA}[1]{{\teal{PA: #1}}} 
\newcommand{\EN}[1]{{\red{EN: #1}}} 

\renewcommand\PA[1]{}
\renewcommand\EN[1]{}

\title{\boldmath 
Lifetimes of light stringy states
}

\preprint{UWThPh 2921-27}

\author{Pascal Anastasopoulos}
\author{and Elias Niederwieser}

\affiliation{Mathematical Physics Group, Department of Physics, University of Vienna, Boltzmanngasse 5 1090 Vienna, Austria}

\emailAdd{pascal.anastasopoulos@univie.ac.at}
\emailAdd{a1306455@unet.univie.ac.at}

\keywords{Light stringy states, Intersecting D-branes, Lifetimes, Dark Matter}
\abstract{In this paper, we evaluate the lifetime of the light stringy states that emerge in intersecting D-brane realisations of the Standard Model. For concreteness, we focus on a light massive scalar decaying into two massless fermions. Given the present experimental lower bounds of the string scale, we extrapolate the lifetime as a function of the intersection angles to very small angles in order to have states with lifetimes of the order of the universe. That provides an alluring example of a massive, very weakly interacting field with a huge lifetime, proposing itself as a potential dark matter candidate.
}

\begin{document}
\maketitle
\flushbottom

\section{Introduction}
\label{sec:introduction}
D-brane model building offers a very rich framework within which realistic realisations of the Standard Model (SM) and beyond are feasible. In this context, the higher dimensional D-branes  are spread within a ten dimensional spacetime covering the four dimensional Minkowski space $\mathcal{M}^{1,3}$ and wrapping cycles in a six dimensional internal Calabi-Yau threefold $CY_3$. 
Strings with both ends on the same stack of D-branes describe the gauge group, while chiral matter appears at the intersections in the internal space of different cycles wrapped by the D-brane stacks \cite{Blumenhagen:2005mu,2005,2004,Kiritsis:2003mc,MarchesanoBuznego:2003axu}.

One of the most exciting properties of these models is that they allow a low string scale \cite{ArkaniHamed:1998rs, Antoniadis:1997zg, Antoniadis:1998ig} even at a few \SI{}{\tera \eV} range~\footnote{At the moment, the lower bound of the string scale is at $8.2-10$ \SI{}{\tera \eV} \cite{Chatrchyan:2011ns, ATLAS:2012pu, Chatrchyan:2013qha, osti_20705761}.}. Therefore, phenomenological studies of these models are particularly interesting and several directions have been extensively analysed, like anomalous $Z'$ physics (see, e.g. \cite{Kiritsis:2002aj, Antoniadis:2002cs, Ghilencea:2002da, Anastasopoulos:2003aj, Anastasopoulos:2004ga, Burikham:2004su, Coriano':2005js, Anastasopoulos:2005ba, Anastasopoulos:2006cz, Anastasopoulos:2008jt, Armillis:2008vp, Fucito:2008ai,  Anchordoqui:2011ag, Anchordoqui:2011eg, Anchordoqui:2012wt}), 
Kaluza-Klein (KK) states (see, e.g. \cite{Dudas:1999gz, Accomando:1999sj, Cullen:2000ef, Burgess:2004yq,       Chialva:2005gt, Cicoli:2011yy, Chialva:2012rq}),
 and purely stringy signatures (see, e.g. \cite{Bianchi:2006nf, Anchordoqui:2007da, Anchordoqui:2008ac, Lust:2008qc, Anchordoqui:2008di, Anchordoqui:2009ja, Lust:2009pz, Anchordoqui:2009mm,  Anchordoqui:2010zs, Feng:2010yx, Dong:2010jt, Carmi:2011dt, Hashi:2012ka, Anchordoqui:2014wha})\footnote{For recent reviews, see \cite{Lust:2013koa, Berenstein:2014wva}.}.

\begin{figure}[t]
\begin{center}
\epsfig{file=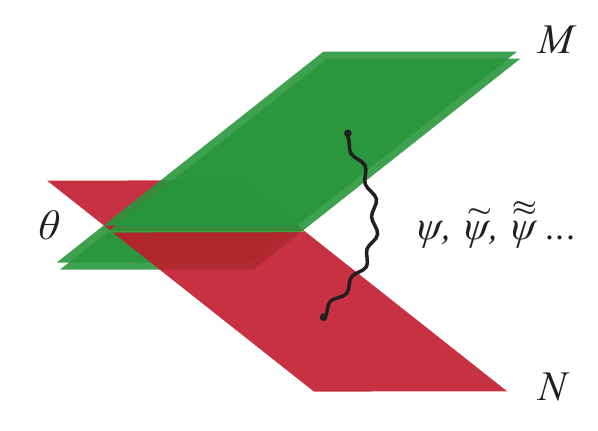,width=60mm}
\caption{Near the intersection of the two stacks of $D_a$-branes and $D_b$-branes open oriented strings are stretched. These open strings transform in the bi-fundamental representation $(N, \overline{M})$ of the gauge group $U(N)\times U(M)$. The chiral matter content of the Standard Model arises from these strings. In fact, a whole tower of SM fields arises, which are identical in all quantum numbers and differ only in mass.}
\label{strings at intersections}
\end{center}
\end{figure}

In this paper, we will take a different direction that has not been studied much and we will focus on the so-called {\it light stringy states} and their empirical implications \cite{Anastasopoulos:2011gn, Anastasopoulos:2013sta, Anastasopoulos:2014lpa, Anastasopoulos:2016yjs}. On such semi-realistic constructions, which consist of intersecting D-branes, we find a whole tower of states living on each intersection, sharing the same quantum numbers and masses, which are proportional to 
\bea
M^2 ~ \sim ~ \theta M_s^2 ~,
\eea
where $\theta$ denotes the intersection angle and $M_s$ characterises the string scale. In this framework, the Standard Model matter content is described by the first massless modes (for reviews see \cite{Blumenhagen:2005mu, Blumenhagen:2006ci, Marchesano:2007de, Cvetic:2011vz} and references therein) and a whole tower of massive states with the same quantum numbers. This is schematically depicted in Fig.~\ref{strings at intersections}. 

Considering this aspect, we focus on the lifetime of light stringy states. Since the decay rates of such states are known, we can proceed and evaluate their lifetimes. It is worth mentioning that these lifetimes have no upper bound. For specific demands on the exact D-brane configuration, the resulting spectrum of possible lifetimes ranges from a few seconds to billions of years and can therefore be considered approximately stable in the latter case.

As an exemplification, we can consider the excited Higgs field. Its mass includes two sources, the vibration of the excited string and some potential affiliated to the untwisted Higgs. The twisted Higgs decays to untwisted SM fields and the decay rate depends on the angles of the intersecting D-branes where the corresponding Higgs state lives. It turns out that the decay rate can be vanishingly small if the intersecting D-branes are almost parallel to each other. It can be shown that in this case its lifetime can be of the order of the lifetime of the universe, which, in accordance to Hubble's Law, is of the order of \SI{e-14}{\s}. \cite{a1,a2,a3,a4}. That provides an alluring example of a massive, very weakly interacting field with a huge lifetime, proposing itself as a potential dark matter candidate.

The paper is organised as follows: In Section~\ref{sec:States}, we discuss the local configuration of three intersecting D-brane stacks. Further, we analyse the states localised at such intersection and  display their corresponding masses and vertex operators. In order to achieve the desired result, we will consider the ground states in the Neveu-Schwarz (NS) and Ramond (R) sectors. In addition, we examine the first excited states in the NS sector. The Yukawas for twisted and untwisted fields located on intersections where the SM is realised gets presented in Section~\ref{Yukawa coupling of untwisted fields}. Finally, we evaluate the lifetime of the twisted states in Section~\ref{Lifetimes of twisted scalars}. There, we also use some assumptions and approximations to set some bounds on these lifetimes. 

\section{States and vertex operators at intersections}\label{sec:States}

\begin{figure}[t]
\begin{center}
\epsfig{file=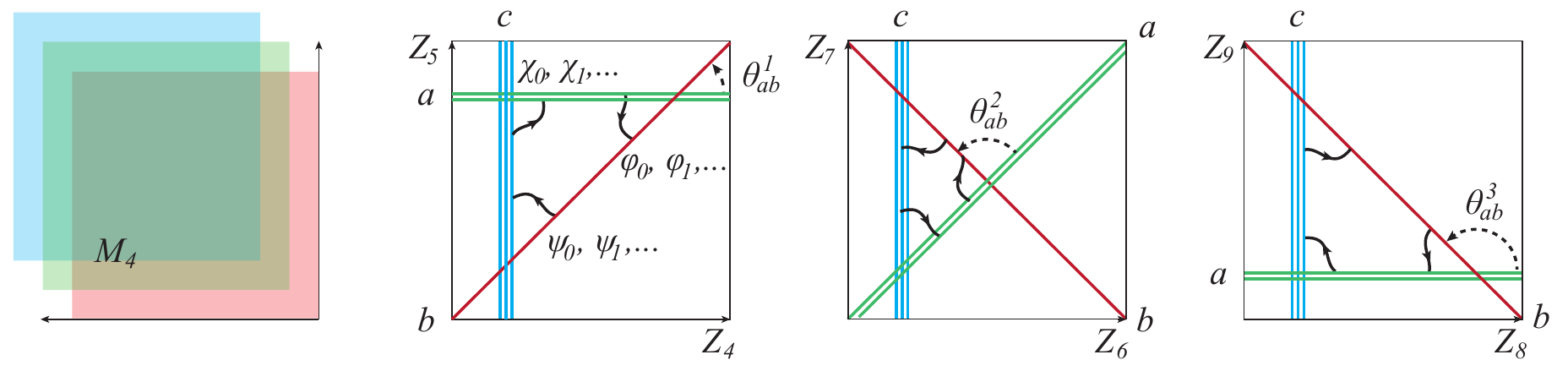,width=150mm}
\caption{A pictorial representation of our setup. While the D-branes fill $\mathcal{M}^{1,3}$, they wrap $3$-cycles in the internal compact $\mathbb{T}^6$, which can be factorised into three $2$-tori $\mathbb{T}^2$. The D-branes can therefore be represented by $1$-cycles which necessarily intersect at non-trivial angles. }\label{D-brane setup}
\end{center}
\end{figure}

In order to perform an explicit calculation, we need to specify the details of the considered setup. The D-brane construction is based on three different stacks of D-branes. More precisely, it consists of  a $D_a$-brane, a $D_b$-brane and a $D_c$-brane, which are wrapped and intersect each other non-trivially on a factorisable $6$-torus $\mathbb{T}^6=\mathbb{T}^2 \times \mathbb{T}^2 \times \mathbb{T}^2$. This configuration for the construction of our D-brane model is illustrated in Fig.~\ref{D-brane setup}. Such a D-brane model gives rise to the following intersection angles $\theta_i$, with $a_i \equiv\frac{\theta}{\pi}$:
\bea
&&a^1_{ab}  > 0,  \qquad \qquad  a^2_{ab}  >0,  \qquad \qquad  a^3_{ab}  < 0,  \qquad \qquad \sum a^i_{ab} =0 ~.\nn \\
&&a^1_{bc}  > 0,  \qquad \qquad  a^2_{bc}  > 0,  \qquad \qquad  a^3_{bc}  <0,  \qquad \qquad ~ \sum a^i_{bc} =0 ~. \label{eq choice of angles} \nn\\
&&a^1_{ca}  < 0,  \qquad \qquad  a^2_{ca}  < 0,  \qquad \qquad  a^3_{ca}  < 0,  \qquad \qquad \sum a^i_{ca} =-2 ~.
\eea
At each intersection there appears a massless fermion which in case of preserved supersymmetry is accompanied by a massless scalar, corresponding to a four dimensional superpartner. In order to guarantee and to provide $\mathcal{N}=1$ supersymmetry, the angles have to satisfy the triangle relations 
\bea
&& a^1_{ab} + a^1_{bc} +a^1_{ca} =0 ~~,~~~~  
a^2_{ab} + a^2_{bc} +a^2_{ca} =0 ~~,~~~~
a^3_{ab} + a^3_{bc} +a^3_{ca} =-2  ~~.~~~~
\label{concrete setup 2}
\eea
Furthermore, we find not only massless matter at each intersection, but also an entire tower of massive copies, whose mass scales with the intersection angle. These excitations are referred to as light stringy states. In scenarios with a low string tension and small intersection angles such states can be fairly light and potentially observed at the Large Hadron Collider (LHC) or future experiments. 

Here, we present the form of the first scalar and fermionic excitations, the VO's and their masses \cite{Anastasopoulos:2016cmg}.

\subsection*{Scalars at angles}
In the following, we focus on the intersections that take place in the $ab$ sector between the $D_a$-brane and the $D_b$-brane. The angles have to satisfy the conventions from (\ref{eq choice of angles}). In particular, this means that two intersection angles are positive, while the last one takes a negative value. The NS vacuum consists of a single massless state, which reads
\bea
\Phi(k)~:~~~~~
\psi_{-\frac{1}{2}-a^3_{ab}} \;|a^1_{ab},a^2_{ab},a^3_{ab}\rangle_{\mathrm{NS}},\quad \text{with} \quad \alpha^{\prime} m^{2}=\frac{1}{2}\sum_{i}^{3} a_{a b}^{i}=0  ~.
\eea
The associated mass squared operator vanishes in that case.
The vertex operator (VO) of this massless state in the canonical $(-1)$ super-ghost picture is given by
\bea
 V_{\Phi}^{(-1)}=g_{\Phi}\left[\Lambda_{a b}\right]_{\alpha}^{\beta} \Phi(x) e^{-\phi} \prod_{I=1}^{2} \sigma^{+}_{a_{a b}^{I}} e^{i a_{a b}^{I} H^{I}} \sigma^{+}_{1+a_{a b}^{3}} e^{i\left(1+a_{a b}^{3}\right) H^{3}} e^{i k X} ~,
\eea
where for the internal space $\mathbb{T}^{6}$ we get contributions from the bosonic twist fields $\sigma^{+}_{a}$ and the bosonised fermionic twist fields $e^{ia_I H_I}$. These twist fields incorporate the mixed boundary conditions of the open string stretched between intersecting branes. The additional $e^{ikX}$ comes from the four dimensional spacetime structure, where the string can move freely. The  Chan-Paton factors $[\Lambda_{ab}]$ indicate that the oriented open string is stretched between the two D-brane stacks $a$ and $b$. On each stack of D-branes there lives a gauge group, thus the indices $\a$ and $\b$ run from one to the dimension of the fundamental representation of that gauge group. The string vertex coupling is denoted by $g_{\Phi}$. 

BRST symmetry requires that a Vertex operator has to obey the physical state condition $[Q_{BRST},V]=0$. Fulfilling this condition gives a double pole which vanishes for $\a' k^2=0$. The form of the BRST charge $Q_{BRST}$ is given in the Appendix~\ref{app:BRSTcharge}. 

Assuming that the angle  $a^1_{ab}$ is smaller than the rest, the lightest stringy states with masses $\a'\,m^2=a^1_{ab}$ include
\bea
&&\widetilde \Phi_1(k)~:~~~~~~~~~~~~~~~~~~
a_{a^1_{ab}} \psi_{-\frac{1}{2}-a^3_{ab}} |a^1_{ab},a^2_{ab},a^3_{ab}\rangle_{\mathrm{NS}}~,  ~~~~~~~  \\
&&\widetilde \Phi_2(k)~:~~~~~~~~~~~~~~~~~~~~~~~
\psi_{-\frac{1}{2}+a^2_{ab}} |a^1_{ab},a^2_{ab},a^3_{ab}\rangle_{\mathrm{NS}} ~.
\eea
The VO for these states are
\bea
V^{(-1)}_{\widetilde \Phi_1} &= g_\Phi [\Lambda_{ab}]^{\beta}_{\a} \, \widetilde \Phi_1(x) \, e^{-\phi} \;\tau_{a^1_{ab}} \,e^{i a^1_{ab} H_1}\; \sigma_{a^2_{ab}} \,e^{i a^2_{ab} H_2} 
 \;\sigma_{1+a^3_{ab}} \,e^{i \left(1+a^3_{ab}\right) H_3}\,e^{ipX} ~, \label{phi1}\\
V^{(-1)}_{\widetilde \Phi_2} &= g_\Phi [\Lambda_{ab}]^{\beta}_{\a} \, \widetilde \Phi_2(x) \, e^{-\phi}\; \sigma_{a^1_{ab}} \,e^{i a^1_{ab} H_1} 
 \;\sigma_{a^2_{ab}} \,e^{-i \left(1-a^2_{ab}\right) H_2}
\; \sigma_{1+a^3_{ab}} \,e^{i a^3_{ab} H_3}\, 
e^{ipX} ~.
\eea
Considering the BRST invariance of the VO's, a double pole appears which vanishes if
\bea
\a' p^2 + a^1_{ab}=0
\label{eq:lightscalarmass}
\eea
for both VO's. Equation~(\ref{eq:lightscalarmass}) confirms that $\tilde\Phi_1$ and  $\tilde\Phi_2$ are massive scalars with mass square $a^1_{ab}/\a'$.
Here, we should notice that the single pole vanishes for both VO's.

Notice that this is not the only mass source for this state. A potential, similar to the one that gives mass to the untwisted state is expected \cite{Anastasopoulos:2009mr}.

\subsection*{Fermions at angles}

The other two states which are involved in our computations are two massless fermions arising from the Ramond sector. These two states are located at the intersections of the $D_b$-brane and $D_c$-brane as well as $D_c$-brane and $D_a$-brane. The two ground states are
\bea
\psi(k) ~:~~~~~
|\,a^1_{bc},a^2_{bc},a^3_{bc}\rangle_{\mathrm{R}} \label{eq vo phi}
~~~~~~~~~\text{and}~~~~~~~~~~ 
\chi(k) ~:~~~~~
|\,a^1_{ca},a^2_{ca},a^3_{ca}\rangle_{\mathrm{R}} \label{eq vo psi} ~.
\eea
Their associated VO's in the canonical $(-1/2)$ super-ghost picture are
\bea
V^{(-\frac{1}{2})}_{\psi} &=& g_{\psi}[\Lambda_{bc}]^\beta_\gamma ~ \psi^{\alpha}_i \, e^{-\phi/2} S_{\alpha}  
\label{VOpsi}\\
&& \times\;\sigma_{a^1_{bc}} \,e^{i \left(a^1_{bc} -\frac{1}{2}\right) H_1} \;
\; \sigma_{a^2_{bc}} \,e^{i \left(a^2_{bc} -\frac{1}{2}\right) H_2} \;
\; \sigma_{1+a^3_{bc}} \,e^{i \left(a^3_{bc} +\frac{1}{2}\right) H_3} \;
\, e^{ikX} ~,~~~~~~~~\nn\\
V^{(-\frac{1}{2})}_{\chi} &=& g_{\chi} [\Lambda_{ca}]^\a_\beta ~ \chi^{\alpha}_i \, e^{-\phi/2} S_{\alpha}  
\label{VOchi} \\
&& \times\;\sigma_{1+a^1_{ca}} \,e^{i \left(a^1_{ca} +\frac{1}{2}\right) H_1}\; \sigma_{1+a^2_{ca}} \,e^{i \left(a^2_{ca} +\frac{1}{2}\right) H_2}\;\sigma_{1+a^3_{ca}}  e^{i \left(a^3_{ca} +\frac{1}{2}\right) H_3}\;
\, e^{ikX} ~.~~~~~~~~\nn\eea
Apart from the spinor wave functions $\psi^{\alpha}_i$ and $\chi^{\alpha}_i$ we have an additional new type of field $S_\alpha$, which denotes a  $SO(1,3)$ spin field determined by the GSO projection. 

The mass squared operator vanishes for the spacetime fermions $\psi$ and $\chi$. Moreover, $\alpha^{\prime}m^2=0$ is independent of the choice of the angles. The physical state condition $[Q_{B R S T}, V_{\psi,\chi}^{(-1 / 2)}]=0$ yields a double and and simple pole.
\begin{itemize}
    \item  The simple pole vanishes if we demand the equation of motion for a massless Weyl fermion, i.e.
    \begin{equation}
        k^\mu \bar\sigma^{\dot a a}_\mu \psi_a(k)= 0 \quad \text{and} \quad k^\mu \bar\sigma^{\dot a a}_\mu \chi_a(k)= 0 ~.
    \end{equation}
     \item  The double pole vanishes for $\alpha^{\prime}k^2=0$.
\end{itemize}

\section{Yukawa coupling of untwisted fields}\label{Yukawa coupling of untwisted fields}
\begin{figure}[t]
\begin{center}
\epsfig{file=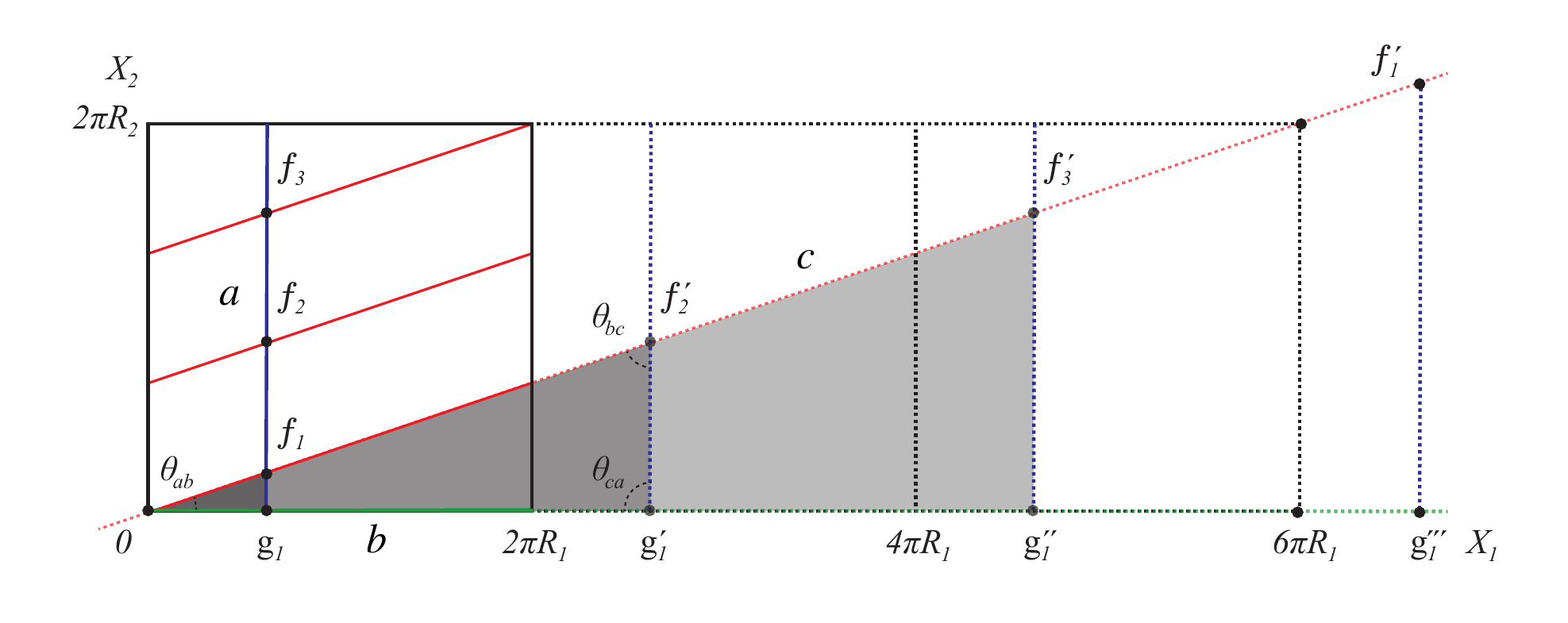, width=1\textwidth}
\caption{Three intersecting D-brane stacks wrapping the cycles $a = (0,1)$, $c= (3,1)$, $b= (1,0)$.}
\label{3triangles}
\end{center}
\end{figure}
The Yukawa couplings of two massless fermions and a massless scalar have been computed carefully in \cite{Cremades:2003qj,Cvetic:2003ch, Abel:2003vv}. According to our D-brane configuration, we adapt it and receive
\begin{equation}
\left|Y_{\Phi \, \psi \, \chi}\right|= g_{\mathrm{op}}(2 \pi)^{-\frac{3}{4}}\left[\Gamma_{1-a_{a b}^{1}, 1-a_{b c}^{1},-a_{c a}^{1}} \Gamma_{1-a_{a b}^{2}, 1-a_{b c}^{2},-a_{c a}^{2}} \Gamma_{-a_{a b}^{3},-a_{b c}^{3},-a_{c a}^{3}}\right]^{\frac{1}{4}} \prod_{i=1}^{3} e^{-\frac{A_{\phi \psi \chi}^{(i)}}{2 \pi \alpha^{\prime}}}~,
\label{YukawaPhipsichi}\end{equation}
with the usual convention
\begin{equation}
    \Gamma_{a, b, c}\equiv\frac{\Gamma(a) \Gamma(b) \Gamma(c)}{\Gamma(1-a) \Gamma(1-b) \Gamma(1-c)}~.
\end{equation}
The expression for the world-sheet instanton contribution is given by the area $A_{\phi \psi \chi}^{(i)}$ of the triangle defined by the three points $f_{\psi, i}, f_{\chi, i}$ and $f_{\phi, i}$ in the $i$\textsuperscript{th} $2$-torus given by
\begin{equation}
  A_{\phi \psi \chi}^{(i)}=\frac{1}{2}\left|\frac{\sin \pi a_{b c}^{i} \sin \pi a_{c a}^{i}}{ \sin \pi a_{a b}^{i}}\right| \left|f_{\chi \psi, i}\right|^{2} ~,  
\end{equation}
with
\begin{equation}
    f_{\psi\chi,i}=g^i -f^i_I + n_i \widetilde{L}^i_c ~.
\end{equation}
Here, the $g^i$ denote the points where the D-brane stacks $a$ and $c$ intersect in the respective $2$-torus $\mathbb{T}^{i}$. Analogously, the $f^i_I$ denote the intersection points in the respective $2$-tori for the D-brane stacks $b$ and $c$. The length $\widetilde{L}^i_c$ is given by
\begin{align}
\widetilde{L}^i_c = 
\frac{\left|I_{ab}\right| }  {\gcd\left(\left|I_{ab}\right|, \left|I_{bc}\right|,\left|I_{ac}\right|\right)} L^i_c\,\,,
\end{align}
where $L^i_c$ denotes the length of the D-brane in the $i$\textsuperscript{th} $2$-torus. Consequently, for our concrete setup $\widetilde{L}^i_c =  L^i_c$ for all three $2$-tori. 

The schematic representation provided in Fig.~\ref{3triangles} illustrates this with an example to further elucidate the notation. The depiction considers three intersecting D-brane stacks wrapping  $1$-cycles according to the wrapping numbers $a = (0,1)$, $b= (1,0)$ and $c= (3,1)$ on the first $2$-torus. The fundamental domain of the rectangular torus is given by the horizontal and vertical line associated to $(2\pi R_1$, $2\pi R_2$). Their corresponding intersection numbers on that torus are $I_{ba}=1$, $I_{ca}=3$, $I_{cb}=-1$. At the intersection of the $a$ and $b$ branes (zero point), a scalar tower is located, corresponding to $\Phi, \widetilde \Phi, \ldots$.
At $f_1$, $f_2$, $f_3$ of the $bc$ sector we find three different families $\psi_1,~\psi_2,~\psi_3$ and their twisted decendance. At the intersection of the $D_a$-brane and the $D_c$-branes there live the $\chi$, $\widetilde \chi,\ldots$ fermions.

\subsection*{Yukawa coupling of a twisted and two untwisted fields}
The Yukawa couplings associated with the twisted $\widetilde \Phi$ and two massless fermions have been elaborated in \cite{Anastasopoulos:2016yjs, Anastasopoulos:2017tvo,Anastasopoulos:2018sqo} and read
\begin{equation}
    \left|Y_{\widetilde \Phi_1 \, \psi \, \chi}\right|=\frac{\left|Y_{\Phi \, \psi \, \chi}\right|}{\sqrt{a_{a b}^{1}}}\left[\Gamma_{1-a_{a b}^{1}, 1-a_{b c}^{1},-a_{c a}^{1}}\right]^{1 / 2} \sqrt{\frac{2 A_{\phi \psi \chi}^{(1)}}{\pi \alpha^{\prime}}}~.\label{YukawaTWISTEDPhipsichi}
\end{equation}
This Yukawa coupling is non-vanishing for a generic setup and it is proportional to the Yukawa couplings of the untwisted states.

It is worth mentioning that this result distinguishes the D-brane vacua from all KK models where the decay of twisted scalars to massless fermions is impermissible. Twisted scalars are copies of the untwisted field with momentum in the compact dimensions. Therefore, decay to untwisted states would violate momentum conservation in the internal directions of KK models.

\section{Lifetimes of twisted scalars}\label{Lifetimes of twisted scalars}

Taking advantage of the Yukawa coupling of a twisted scalar to two massless fermions and the fact that there are no other decay channels at this order, we can compute the lifetime of such a state. The decay rate for a massive scalar particle ${\widetilde \Phi_1}$ to decay at rest into two massless fermions is given by
\bea
\Gamma = \frac{m_{\widetilde \Phi_1}}{8 \pi} \sum_{ij} \left|Y_{{\widetilde \Phi_1} \psi_i \chi_j}\right|^2 ~.
\eea
The sum runs over flavours and colours. The lifetime of a particle is related to the decay rate according to $\tau=\Gamma^{-1}$. For the twisted field we therefore obtain
\bea
\tau_{\widetilde \Phi_1}=\frac{8 \pi}{m_{\widetilde \Phi_1} \sum_{ij} |Y_{{\widetilde \Phi_1} \psi_i \chi_j}|^2} ~.
\eea
Assuming that the leading contribution to the Yukawa coupling is solely coming from the volume of the first and also smallest triangle, which is formed by the $D_a$-brane, $D_b$-brane and $D_c$-branes, we find
\bea
\tau_{\Phi_1}&=&
\frac{8 \pi}{m_{\Phi_1}}
\frac{1}{|Y_{\Phi \psi  \chi}|^2} ~,
~~~~~\\
\tau_{\widetilde \Phi_1}&=&
\frac{4 \pi^2}{m_{\widetilde \Phi_1}}
\frac{a^1_{ab}}{|Y_{\Phi \psi  \chi}|^2}
\left[\Gamma_{a_{a b}^{1}, a_{b c}^{1},1+a_{c a}^{1}}\right]
\frac { \alpha'}{ A^{(1)}_{\Phi \psi  \chi}} ~.~~~~~
\eea
%
%
%
%
%
%
%
%
Consequently, we can express the lifetime of the twisted field as
\begin{equation}
\frac{\tau_{\tilde\Phi_1}}{\tau_{\Phi_1}}=
\frac{\pi\a' a_{ab}^{1}}{|f_{\psi\chi,1}|^2}
\frac{m_{\Phi_1}}
{m_{\tilde\Phi_1}}
\left[\Gamma_{a_{a b}^{1}, a_{b c}^{1},1+a_{c a}^{1}}\right]
\left|\frac{\sin\pi a^1_{ab}} {\sin\pi a^1_{bc} \sin\pi  (1+a^1_{ac})}\right| ~.
\label{lifetimesDEV3}
\end{equation}

The expression \eqref{lifetimesDEV3} can further be simplified if we assume that the distance $f_{\psi\chi,1} \sim a^1_{ab}\, R$ with $R$ corresponding to the radius of the internal torus is taken of order of the string length $R\sim l_s=2\pi\sqrt{\a'}$. This leads to
\begin{equation}
\tau_{\tilde\Phi_1}\sim
\frac{1 }{4\pi a_{ab}^{1}}
\frac{m_{\Phi_1}}
{m_{\tilde\Phi_1}}
\left[\Gamma_{a_{a b}^{1}, a_{b c}^{1},1+a_{c a}^{1}}\right]
\left|\frac{\sin\pi a^1_{ab}} {\sin\pi a^1_{bc} \sin\pi  (1+a^1_{ac})}\right|\,\tau_{\Phi_1} ~.
\label{lifet1}
\end{equation}
To investigate (\ref{lifet1}) it is useful to further simplify it by using an approximation for small angles. More details can be found in the Appendix~\ref{Gamma}. Demanding that $a^1_{ab}<\epsilon_1,\, a_{b c}^{1}<\epsilon_2$ and $a_{c a}^{1}<\epsilon_3$ with $\epsilon_i \ll 1$ produces
\begin{equation}
\tau_{\tilde\Phi_1}\sim
\frac{1}{4\pi}\frac{m_{\Phi_1}}
{m_{\tilde\Phi_1}}
\left[\frac{1}{a_{a b}^{1}(a_{b c}^{1})^2}+\frac{2\gamma a_{c a}^{1}}{a_{a b}^{1}(a_{b c}^{1})^2}+\frac{4\gamma^{2}}{a^1_{bc}}\right]
\tau_{\Phi_1}  ~.
\label{lifetimesDEV1}
\end{equation}
Moreover, by assuming that $a^1_{ab}, a^1_{bc}, a^1_{bc} \sim \varepsilon$ where $\varepsilon \rightarrow 0$ we deduce that
\begin{equation}
    \tau_{\tilde\Phi_1}\sim\frac{1}{4\pi}\frac{m_{\Phi_1}}
{m_{\tilde\Phi_1}}
\left[\frac{1}{\varepsilon^3}+\frac{2\gamma }{\varepsilon^2}+\frac{4\gamma^{2}}{\varepsilon}\right]
\tau_{\Phi_1}
\label{lifetimeAfterManipulations}\end{equation}
The lifetime depends on the values of the angles $a^1_{ab}, a^1_{bc}$. It is clear that the lifetime of the twisted scalar can be very long for very small angles. For example, for almost parallel $a$ and $b$ branes this lifetime
can be of order of the age of the universe.

\subsection{Lifetime of the twisted Higgs at D-brane realisations of the SM}

Here, we apply our results on some D-brane realisations of the SM. In such a framework, the SM fields correspond to open strings, attached on different stacks of D-branes. The gauge fields and the emerging Yang-Mills theories at low energies are related to strings with both ends beginning and terminating on the same stack of D-branes. To ensure such a semi-realistic model, we introduce $a)$ a stack of three branes (so-called \textit{colour} or \textit{baryonic} branes), which give rise to the gluons, $b)$ a stack of two branes (\textit{left} branes) realising the $SU(2)_L$ gauge fields and $c)$ some single D-branes (\textit{right} brane). The hypercharge is obtained by a linear combination of abelian factors living on each stack. On the other hand, the chiral matter fields live on D-brane intersections. For example, the left-handed quark doublet is associated to a string with one end on the colour brane and the other end on the left brane \cite{Aldazabal:1999tw, Antoniadis:2000ena, Antoniadis:2002qm, Anastasopoulos:2006da, Ibanez:2012zz}.

In order to apply our results on a D-brane realisation of the SM, we assume that the stack $c$ consists of colour branes. In the case of the $a$ brane, on the other hand, we have a stack of two coincident left branes. The stack $b$ completes the triangle and corresponds to a right brane (see Fig.~\ref{D-brane setup}). In these intersections the oriented open strings receive their $SU(3)_C$, $SU(2)_L$ and $U(1)_R$ charge from the respective branes, resulting in states living in the intersections $ca$/$cb$/$ba$, which correspond to the SM fields $Q_L$/$d_R$/$H_u$ respectively \footnote{D-brane realisations of the SM are typically supersymmetric with two Higgses $H_u$ and $H_d$. However, we focus on $H_u$ and similar discussions can be applied for the $H_d$.}\footnote{The hypercharge of the model is given by $Q_Y=\frac{1}{6}Q_c+\frac{1}{2}Q_b.$}. The Yukawa coupling that gives masses to the top quark is given by \eqref{YukawaPhipsichi} \footnote{Notice that in many D-brane realisations of the SM the Yukawas are not always present and D-instantons contribute in order to provide the missing mass terms \cite{Anastasopoulos:2009mr, Cvetic:2009yh}.}. 
Moreover, at the intersection $ab$ we find the excited Higgs field $\tilde H_u$ and the Yukawa coupling associated to $\bar Q_L d_R \tilde H_u$ represented by \eqref{YukawaTWISTEDPhipsichi}. 

In such generic case, the life time of the $\tilde H$ is given by \eqref{lifetimeAfterManipulations} where the scalar $\Phi$ is the Higgs boson $H$, and the $\tilde\Phi$ is the first excitation $\tilde H$. In that case,  $m_{\Phi}\simeq \SI{125}{\giga \eV}$ and $\tau_{\Phi} \simeq \SI{e-22}{\s}$. Assuming that the mass of the excited Higgs is $m_{\tilde\Phi}\ge m_{ \Phi}$ \footnote{The twisted Higgs has a second source of mass term. A potential, similar to the untwisted Higgs that provides a mass term that we assume to be at least of the order $M_{H}$.} and the string scale is around a few \SI{}{\tera \eV}, we can estimate the values for the angle $a^1_{ab}\sim \SI{e-14}{}$. Obviously, this is a very tiny value, where the branes $a$ and $b$ are almost parallel, but still it shows that such twisted states can be very long-lived at extreme cases. Therefore, such a massive, long-lived particle could be a dark-matter candidate. 
However, we need to perform a detailed analysis of such a model and its implications on the cosmological history (early matter domination, entropy injection, etc.) in order to determine its viability \cite{AMN}.



\vspace*{0.3cm}

\section*{Acknowledgements}
\label{ACKNOWL}
\addcontentsline{toc}{section}{Acknowledgements}

We would like to thank Massimo Bianchi, Dario Consoli and Yann Mambrini for informative and enlightening discussions. E.N. would also like to thank Inge Sader and Jasmin Khlifi, for their hospitality during the final stages of this work.
P.A. was supported by FWF Austrian Science Fund via the SAP P30531-N27.

\vspace*{0.3cm}
\begin{appendix}
\section*{Appendix}
\label{Appendix}
\section{BRST charge and physical states}
\label{app:BRSTcharge}
The invariance of physical VO's under the BRST charge (presenting only the matter part)
\bea
Q_{BRST}&=& \oint \frac{dz}{2 \pi i}\bigg\{ e^{\phi} \frac{ \eta}{\sqrt{2 \alpha'}} \bigg( i\partial X^{\mu} \, \psi_{\mu} + \sum^3_{I=1} \partial Z^{I } \, e^{-iH_I} + \sum^3_{I=1} \partial \ov Z^{I } \, e^{iH_I}\bigg) \label{eq BRST charge}\\
&&~~~~+ \frac{c}{\a'} \bigg(i\partial X^\mu i \partial X_\mu - \frac{\a'}{2} \psi^\mu \partial \psi_\mu + \sum^3_{I=1} \Big(\partial Z^I \partial \ov Z_I - \frac{\a'}{2} e^{-iH_I} \partial e^{iH_I} \Big)\bigg) \bigg\} + ...\nn
\eea
denotes that 
\bea
[Q_{BRST},V]=0 
\label{BRSTinvariance}
\eea
where $V$ is any physical vertex operator. Using the operator product expansions (OPE's) we get a double and a simple pole which should vanish. 
\section{State - Vertex operator dictionary \label{app dictionary} }

In the following table  
\begin{eqnarray*}
\begin{tabular}{llll} 
\toprule
\multicolumn{2}{l}{Positive angles}                                                                                                                                                      & \multicolumn{2}{l}{Negative angles}                                                                                                                                                     \\ 
\hline
state                                          & vertex operator  & state                                        & vertex operator   \\ 

$ | \, a \, \rangle_R $                        & $ e^{i \left(a-\frac{1}{2}\right) H(z)} \sigma^+_{a}(z) $                                                                               & $ | \, a \, \rangle_R $                      & $ e^{i \left(\frac{1}{2}+a\right) H(z) } \sigma^-_{-a}(z) $                                                                              \\
$ \alpha_{-a} | \, a\, \rangle_R $             & $ e^{i \left(a-\frac{1}{2}\right) H(z) }\tau^+_{a}(z) $                                                                                 & $ \alpha_{a} | \, a\, \rangle_R $            & $ e^{i \left(\frac{1}{2}+a\right) H(z) }\widetilde\tau^-_{-a}(z) $                                                                       \\
$ \psi_{- a}| \, a \, \rangle_R $              & $ e^{i \left(a+\frac{1}{2}\right)H(z)} \sigma^+_{a}(z) $                                                                                & $ \psi_{ a}| \, a \, \rangle_R $             & $ e^{i \left(a +\frac{1}{2}\right)H(z)} \sigma^-_{-a}(z) $                                                                               \\
$ \alpha_{-a}\, \psi_{-a} | \, a\, \rangle_R $ & $ e^{i \left(a +\frac{1}{2}\right) H(z) }\tau^+_{a}(z) $                                                                                & $ \alpha_{a}\, \psi_{a} | \, a\, \rangle_R $ & $ e^{i \left(a +\frac{1}{2}\right)H(z) }\widetilde\tau^-_{-a}(z) $                                                                       \\
\bottomrule
\end{tabular}
\end{eqnarray*}
we display the dictionary between state and their corresponding vertex operators in the Ramond sector, which we apply in Section~\ref{sec:States} to determine the vertex operators of the massless boson $\Phi$ and the bosonic light stringy state $\widetilde{\Phi}$. For more details, see~\cite{Anastasopoulos:2011hj}.
\section{Gamma limits}
\label{Gamma}

Expanding the Gamma function around $u=0$ yields 
\begin{equation}
\Gamma(u)=\frac{1}{u}(1-\gamma u) 
+\frac{1}{12}(6\gamma^2+\pi^2)
+{\mathcal{O}}(a) \quad \text{and} \quad
\Gamma(1-u)=1+\gamma u +{\cal O}(u^2)~,
\end{equation}
with an error lower than $0.1\,\%$ for any $0<u\leq \frac{1}{10}$ and $\gamma$ denoting the \textit{Euler-Mascheroni} constant. Further, we can use this approximation to determine
\begin{equation}
\frac{\Gamma(u)}{\Gamma(1-u)}=\frac{1}{u}(1-2\gamma u + 2\gamma^2 u^2) +{\cal O}(u^2) \quad \text{and} \quad \frac{\Gamma(1+u)}{\Gamma(-u)}=-u+2\gamma u^2 +{\cal O}(u^3)~.
\end{equation}
\end{appendix}

\bibliographystyle{JHEP}

\end{document}